\documentclass[12pt]{article}
\usepackage{amsmath,amssymb,geometry}
\usepackage{authblk} 
\usepackage{graphicx}
\usepackage[round,authoryear]{natbib} 
\title{Three Lenses on the AI Revolution: \\Risk, Transformation, Continuity}

\author{Masoud Makrehchi\thanks{Corresponding author: \texttt{<masoud.makrehchi@ontariotechu.ca>}}}
\affil{Department of Electrical, Computer \& Software Engineering,\\
Ontario Tech University, Oshawa, Ontario, Canada}
\date{}
\begin{document}
\maketitle


\begin{abstract}
Artificial Intelligence (AI) has emerged as both a continuation of historical technological revolutions and a potential rupture with them. This paper argues that AI must be viewed simultaneously through three lenses: \textit{risk}, where it resembles nuclear technology in its irreversible and global externalities; \textit{transformation}, where it parallels the Industrial Revolution as a general-purpose technology driving productivity and reorganization of labor; and \textit{continuity}, where it extends the fifty-year arc of computing revolutions from personal computing to the internet to mobile. Drawing on historical analogies, we emphasize that no past transition constituted a strict singularity: disruptive shifts eventually became governable through new norms and institutions. 

We examine recurring patterns across revolutions, democratization at the usage layer, concentration at the production layer, falling costs, and deepening personalization, and show how these dynamics are intensifying in the AI era. Sectoral analysis illustrates how accounting, law, education, translation, advertising, and software engineering are being reshaped as routine cognition is commoditized and human value shifts to judgment, trust, and ethical responsibility. At the frontier, the challenge of designing moral AI agents highlights the need for robust guardrails, mechanisms for moral generalization, and governance of emergent multi-agent dynamics. 

We conclude that AI is neither a singular break nor merely incremental progress. It is both evolutionary and revolutionary: predictable in its median effects yet carrying singularity-class tail risks. Good outcomes are not automatic; they require coupling pro-innovation strategies with safety governance, ensuring equitable access, and embedding AI within a human order of responsibility.
\end{abstract}

\section{Introduction}

A central question in contemporary debates on artificial intelligence (AI) is whether its trajectory fits within the established pattern of technological revolutions or represents a genuine singularity. This framing is crucial because it determines how societies anticipate governance challenges, design institutions, and prepare for potential risks. If AI follows a ``normal science'' path, it may resemble electricity or computing, where steady progress, diffusion, and regulation gradually take hold. If, however, it marks a singularity, then familiar laws, trends, and governance tools could quickly become obsolete.


The first perspective treats AI as a continuation of historical general-purpose technologies such as steam engines, electricity, and computing. According to this view, progress is largely path-dependent: capabilities improve in tandem with advances in data availability, algorithmic sophistication, and computational power \citep{mokyr1990lever,gordon2016rise}. 

Institutions, though often slow, eventually adapt. Markets develop new products and business models, standards are codified, and regulatory frameworks catch up to mitigate externalities. A central feature of this trajectory is the \textit{rebundling of work}. Tasks migrate from humans to machines, and jobs reorganize around new leverage points such as human judgment, oversight, safety, and system integration. While externalities exist, including energy consumption, bias, misinformation, and safety incidents, they are regarded as serious but ultimately manageable through audits, incentives, and legal mechanisms \citep{brynjolfsson2017machine}.

The alternative hypothesis views AI as a discontinuity, a transformative rupture rather than an incremental advance. In this framing, AI systems could improve themselves in feedback loops, accelerating capabilities beyond human control. Governance surfaces become unreliable: capabilities diffuse so quickly that traditional oversight tools, standards, regulation, even international agreements, fail to constrain behavior \citep{bostrom2014superintelligence,russell2019human}.

A qualitative shift in agency is another concern. Autonomous agents may begin to plan, act, and coordinate in open environments, raising alignment problems with human goals. Finally, singularity implies a macroeconomic break: productivity and wealth could concentrate at unprecedented rates, destabilizing social contracts and political institutions.


Empirical evidence so far provides signals for both continuity and discontinuity. On one hand, scaling laws continue to hold, predicting model performance with reasonable accuracy. Gains still track investments, and most economic value accrues in familiar verticals such as coding, customer support, analytics, and education. Cost curves and engineering constraints remain binding. These are strong signals of continuity \citep{stanford2025aiindex}.

On the other hand, discontinuity cannot be dismissed. Large models have shown surprising generalization across tasks, novel forms of tool use, and rapid diffusion via open-source releases. The pace of iteration in the ecosystem is unprecedented, and safety incidents can now propagate globally in hours rather than years. These dynamics suggest that singularity-class risks, though not dominant in everyday practice, remain on the table.

To track regime shifts, five practical tests can be applied:
\begin{enumerate}
    \item \textbf{Forecastability Test:} If simple scaling models continue to predict performance, AI remains within the normal-science regime. Abrupt failures across domains would support the singularity view.
    \item \textbf{Self-Improvement Loop Test:} Evidence that systems materially improve their own training, evaluation, or deployment without human oversight would indicate discontinuity.
    \item \textbf{Governance Tractability Test:} If audits, red-teaming, and incident reporting reliably reduce harms, governance remains tractable. If harm scales faster than oversight, the singularity dynamic strengthens.
    \item \textbf{Resource-Constraint Test:} As long as compute, energy, and data remain limiting factors, AI stays bounded. If synthetic data and abundant compute remove bottlenecks, regime change is plausible.
    \item \textbf{Socioeconomic Absorption Test:} If labor markets, education systems, and law adapt within years, continuity prevails. If they are overwhelmed within months, discontinuity becomes more likely.
\end{enumerate}


The right framing is not to ask ``Which is it?'' but ``Where are we on both curves?'' Most uses of AI today, from coding assistants to customer service automation, fit the normal-science trajectory and can be governed with established tools such as standards, audits, liability, and best practices. Extreme applications, however, call for a tail-risk playbook, including capability thresholds, independent audits, staged deployment, kill-switches, and international coordination.

For practitioners and policymakers, the immediate implications are threefold: (1) AI will continue to deflate the cost of cognition, bringing genuine productivity gains; (2) homogenization and error risks require verification-first practices; and (3) governance is not optional but the essential technology that makes AI safe to deploy at scale.


On the other side, history offers no examples of true singularities. What we observe instead are \textit{phase transitions}, rapid but still governable shifts. Events that come closest outlined in Table \ref{tab1}.

\begin{table}[h!]
\centering
\label{tab1}
\caption{Historical Events and Their Characteristics}
\begin{tabular}{|p{4cm}|p{3cm}|p{8cm}|}
\hline
\textbf{Event} & \textbf{Date} & \textbf{Description} \\
\hline
Invention of Writing & c.~3200~BCE & Enabled external memory, administration, contracts, and science, but diffusion occurred slowly over centuries. \\
\hline
Printing Press & 1450s & Collapsed the cost of copying ideas, catalyzing the Reformation and scientific communities; diffusion and governance took decades. \\
\hline
Black Death & 1347--1351 & Triggered labor scarcity and social reordering, but did not permanently alter governance systems. \\
\hline
Columbian Exchange & After~1492 & Integrated global biological and economic systems, producing massive demographic shifts, yet absorbed within evolving state and market dynamics. \\
\hline
Industrial Revolution & 18th--19th~century & Broke pre-modern growth ceilings through fossil fuels and mechanization, but unfolded in waves and remained bounded by political and resource constraints. \\
\hline
Nuclear Weapons & From~1945 & Introduced civilization-scale tail risk; however, median life continued under governance mechanisms such as deterrence and treaties. \\
\hline
The Internet & 1990s--2000s & Reduced communication costs nearly to zero, reshaping media, commerce, and politics, yet constrained by economics, law, and human attention. \\
\hline
\end{tabular}
\end{table}

The lesson is that while discontinuities in growth or risk occur, they are historically absorbed into new institutional frameworks. AI will likely follow this dual pattern: governable in its median effects, but potentially singular in its tail risks. Planning for both is therefore essential.

The remainder of this paper is organized as follows. Section 2 introduces the three interpretive lenses, risk, transformation, and continuity, that frame AI’s position in the broader arc of technological change. Section 3 draws lessons from the Industrial Revolution to contextualize AI’s economic and institutional effects. Section 4 compares four technological revolutions of the past fifty years, highlighting common structural patterns and the unprecedented adoption speed of AI. Section 5 examines how AI reshapes perception, knowledge production, and cultural value, while Section 6 explores its implications for writing, reading, and communication in an age of interactive, adaptive text. Finally, Section 7 concludes by synthesizing these perspectives and outlining the dual challenge of harnessing AI’s transformative potential while governing its systemic risks.

\section{Three Lenses on the AI Revolution}

Artificial Intelligence can be interpreted through three complementary but distinct lenses: \textit{risk}, \textit{transformation}, and \textit{continuity}. Each captures different dimensions of AI’s impact on society, economy, and governance. Importantly, these lenses are not mutually exclusive; they reflect simultaneous truths about a technology that is both evolutionary and disruptive.

\subsection{Lens A: Risk (AI $\approx$ Nuclear)}

The first perspective emphasizes AI’s resemblance to nuclear technologies in terms of consequences rather than physics. Both domains carry irreversible risks and tail outcomes where failures may be catastrophic. For AI, one-way doors include runaway automation, autonomous agents acting without alignment, or the weaponization of generative capabilities for biosecurity or cyber-offense. 

Like nuclear technology, AI is inherently \textit{dual-use}: the same models that can generate scientific discoveries or assist in drug design may also enable malicious applications, such as automated cyberattacks or novel pathogen synthesis. The proliferation pressure is particularly acute: while nuclear fission requires rare materials and complex infrastructure, AI models spread cheaply once weights are released or know-how diffuses into open communities \citep{brundage2018malicious}.

Another risk dimension is opacity. Modern AI systems are often black boxes with emergent behaviors, producing \textit{unknown unknowns} in causal reasoning and decision-making. Governance must therefore prioritize safety cases, third-party audits, incident reporting, compute and accountability tracking, and export controls for sensitive capability bundles. Additional obligations include red-teaming, continuous evaluation, and strong privacy/security baselines \citep{bostrom2014superintelligence}. 

The key difference from nuclear technology lies in diffusion speed. AI has far lower barriers to entry; thus, governance must operate at \textit{software speed} and be globally interoperable. Traditional treaty-like approaches, effective in arms control, may be too slow.

\subsection{Lens B: Transformation (AI $\approx$ Industrial Revolution)}

The second perspective frames AI as a new general-purpose technology (GPT) with parallels to the Industrial Revolution. Just as mechanization and electrification transformed labor and productivity, AI extends automation into the cognitive domain. Productivity gains compound when paired with complementary factors such as data availability, computational power, and organizational redesign \citep{brynjolfsson2017machine}.

AI’s labor-market effects are \textit{skill-biased}: tasks, rather than entire jobs, are automated, shifting demand toward human roles that emphasize judgment, oversight, ethics, and cross-domain integration. Entirely new sectors are emerging, including agentic services, synthetic research platforms, AI safety and evaluation firms, and edge inference ecosystems. 

The transformation, however, is turbulent. As with industrialization, displacement often precedes widespread benefit. Bargaining power, wage distribution, and market concentration shape outcomes more than the technology itself. Effective policy levers include rapid reskilling programs, wage insurance, competition policies at platform layers, open standards, and ensuring compute and connectivity as public infrastructure. Digital and AI literacy become essential components of the new social contract \citep{acemoglu2021ai}. 

The analogy with the Industrial Revolution underscores a central insight: long-run growth is likely, but short-run shocks are inevitable. Outcomes depend less on technological capability and more on the institutional frameworks that govern its deployment.

\subsection{Lens C: Continuity (AI as the Fourth Installment of Tech Revolution)}

A third interpretation places AI within the 50-year arc of computing: from personal computing to the internet, to mobile networks, and now to AI. Each wave extends two axes: \textit{what is automated} and \textit{who can use it}. 

The trajectory shows continuity. PCs automated data processing, the internet democratized access to information, and mobile technology enabled contextual, always-on communication. AI continues this line by automating knowledge processing and decision support. As before, democratization occurs fastest at the user layer, while production and research remain concentrated due to scale economics \citep{gordon2016rise}. 

Market structures exhibit power-law dynamics: a few dominant players capture most of the value, while open-source communities provide a durable counterweight. Cost curves in compute, storage, bandwidth, and tokens continue to decline, enabling new applications. The personalization trend also deepens: from personal computing to personal communication, and now to personal co-pilots that adapt to user goals, preferences, and constraints. 

Yet continuity has its trade-offs. Each wave widened privacy exposure and increased \textit{moral distance}, the gap between an individual’s actions and their consequences at scale \citep{floridi2021unified}. Without safeguards, AI risks intensifying both.

\subsection{Reconciling the Lenses}

At first glance, the three lenses may appear contradictory. In practice, they are complementary: 
\begin{itemize}
    \item Nuclear-like risks demand strong safety governance, including evaluation, containment, provenance, and liability. 
    \item Industrial-like transformation calls for pro-growth complements: skills, infrastructure, competition, and open standards. 
    \item Continuity reminds us to expect concentration at the production layer, democratization at the use layer, persistent open-source countercurrents, and growing privacy and moral-distance challenges. 
\end{itemize}

Together, these perspectives reveal AI as both a high-tail-risk technology, a driver of broad economic transformation, and a continuation of computing’s historical arc.



In sum, the three lenses illuminate AI’s complexity: it is simultaneously risky, transformative, and continuous. The task ahead is to prepare for each dimension, recognizing that governance, economic design, and ethical frameworks must evolve in tandem.

\section{Lessons from the Industrial Revolution for the AI Transition}

The Industrial Revolution of the eighteenth and nineteenth centuries offers a powerful historical analogy for understanding today’s AI transition. Just as mechanization and electrification redefined production and labor, artificial intelligence is poised to redefine cognition and decision-making. By comparing recurring patterns between the two eras, we can anticipate opportunities, risks, and institutional challenges.
\begin{itemize}
    \item Automation and Scale: During the Industrial Revolution, mechanization and assembly lines enabled production at unprecedented scale. Tasks once performed by artisans were absorbed into standardized factory systems. Today, AI plays an analogous role by scaling cognitive tasks. From code assistants to autonomous operations in logistics, finance, and media, model-driven automation multiplies human productivity. Organizations are being pressured to redesign workflows around “AI-first” processes, in the same way that nineteenth-century firms reorganized production around mechanized factories \citep{mokyr1990lever,brynjolfsson2017machine}.

\item Deflationary Pressure on Cost: Industrialization lowered unit costs as fixed capital investments were spread across vast volumes of output. Mass production made goods accessible to broader populations. Similarly, once trained, AI models drive the marginal cost of many cognitive tasks toward zero. Translation, summarization, drafting, image generation, and even software development can be delivered at negligible cost. Price collapses in cognitive services are already reshaping industries, echoing the deflationary dynamics of industrial manufacturing \citep{gordon2016rise}.

\item Standardization of Quality: The nineteenth century also introduced interchangeable parts and process control, which standardized product quality. For AI, guardrails, evaluation suites, and benchmarks now serve an analogous role. Instead of relying solely on human expertise, we are entering an era where “good-enough by default” outputs become standardized through automated evaluation pipelines. Service-level agreements may soon be defined by benchmark performance rather than individual reputation.

\item The Barbell Effect: Commodities and Bespoke Goods: Industrialization hollowed out mid-tier production. Cheap, mass-produced goods dominated markets, while only elite customized items survived at premium prices. A similar barbell effect is emerging in knowledge work. AI is commoditizing mid-tier drafting, analysis, and support services. What remains scarce, and thus highly valuable, are tasks requiring taste, trust, and deep contextual awareness: high-stakes legal advice, strategic consulting, and brand-specific creative work. This bifurcation mirrors how industrial-era consumers shifted between mass-market products and luxury artisanal goods.

\item Production Standards and Consumption Habits: Railway timetables, packaging innovations, and retail formats during industrialization reshaped not only production but also consumption habits. Today, prompt-based interaction, co-pilot design, and conversational user interfaces are reshaping how individuals search, learn, and make decisions. New standards in human-computer interaction may become as culturally pervasive as standardized time was in the nineteenth century.

\item Labor Market Transformation: Industrialization eliminated certain occupations while creating new roles, mechanics, managers, and machine operators. Similarly, AI does not eliminate jobs wholesale but automates tasks within them. Roles will reorganize around human strengths: problem framing, judgment under uncertainty, cross-domain integration, and relationship management. Reskilling and adaptive education are not optional but essential to navigating this transition \citep{acemoglu2021ai}.

\item Power Concentration and Governance:  Factories themselves were not inherently destabilizing; rather, the concentration of ownership and capital generated political and social tensions. For AI, the most significant risks lie in concentrated control over compute, data, and platform infrastructure. Governance must address these bottlenecks through auditing, antitrust vigilance, and open standards, echoing nineteenth-century reforms in labor law and market regulation.

\item Displacement, Migration, and Social Reordering: Industrialization drove urbanization and transformed family structures as people migrated to cities for factory work. In the AI era, digital migration is already underway. Remote work, global labor markets, and credentialing systems are shifting how individuals enter and progress in professional life. Education, licensing, and safety nets must adapt more quickly than institutions did in the nineteenth century.

\item Environmental Externalities:
Industrial growth came with heavy environmental costs: pollution, deforestation, and carbon emissions. AI, too, has significant externalities. Large model training requires vast energy and water resources, while device turnover generates e-waste. Unless sustainability practices are embedded early, carbon accounting, efficiency targets, and life-cycle design, AI risks locking in long-term environmental costs.

\item Moral Distance and Alienation: Finally, industrialization lengthened supply chains, obscuring the connection between consumer choices and distant harms. AI introduces new forms of moral distance: algorithmic mediation blurs responsibility for decisions, amplifies harms at scale, and facilitates manipulation. As synthetic media increasingly competes with lived reality, societies must design countermeasures such as provenance standards, friction for high-risk actions, and norms that prioritize human agency.
\end{itemize}

Several lessons emerge from this historical analogy:
\begin{itemize}
    \item Design for scale but rebundle work around uniquely human capabilities.
    \item Expect the barbell dynamic: compete at the low end with cost efficiency or at the high end with bespoke trust and expertise.
    \item Standardize quality through transparent audits, metrics, and continuous evaluation loops.
    \item Disperse power by encouraging open formats, data portability, and oversight over compute concentration.
    \item Price in externalities from the beginning: carbon-aware scheduling and resource-efficient design are critical.
    \item Reduce moral distance by embedding traceability, accountability, and user rights into AI systems.
    \item Protect reality by privileging provenance, media literacy, and human-centered product defaults.
\end{itemize}

These parallels remind us that AI’s trajectory, like industrialization, is not solely about technological potential but about the institutions and safeguards that shape its integration into society.

\section{Four Technological Revolutions in the Last 50 Years: A Short Analysis}

From the late twentieth century to the present day, societies have experienced four successive technological revolutions. Each of these episodes, while distinct in form, shares a common logic: the automation of a critical human task at scale, and the democratization of that capability to the wider population. The revolutions of personal computing, the internet, mobile networks, and artificial intelligence form a trajectory in which both the scope of automation and the speed of adoption expand over time.
\begin{itemize}
    \item Personal Computing (Late 1970s -- Early 1980s): 

The personal computing revolution shifted computational capacity from centralized government and corporate mainframes to individuals and small businesses. It automated data processing, empowering users to run spreadsheets, word processors, and database applications that previously required large institutional infrastructures. This transition democratized access to computing power, enabling professionals and eventually households to engage with digital tools directly \citep{ceruzzi2003history}.

Penetration remained limited in the early years. By the mid-1990s, only around 10\% of the global population---concentrated in developed countries---owned a personal computer. Adoption speed was relatively slow, taking nearly two decades to reach moderate global diffusion. Nevertheless, the personal computer created the platform on which subsequent revolutions were built.
\item Internet Revolution (Late 1980s -- Early 1990s): The second revolution was catalyzed by the internet. At its core, the internet automated large-scale information generation, sharing, and retrieval. What was once localized or institutionally siloed knowledge became globally accessible. This democratization of information profoundly altered commerce, governance, and culture \citep{castells2000rise}.

By 2018, approximately 50\% of the world’s population was online. Although its adoption trajectory spanned about 25 years, the internet spread faster than personal computing due to network effects and falling infrastructure costs. Its disruptive power lay in reconfiguring industries such as media, retail, and finance, and in reshaping how individuals connect, learn, and collaborate.

\item Mobile and Smartphone Revolution (Late 1990s -- Early 2000s): Mobile networks and smartphones extended the internet’s promise, embedding connectivity into everyday life. The automation here was of personal communication and mobile data access. For the first time, billions of people were continuously connected to information flows, markets, and one another.

This revolution democratized instant connectivity and mobile computing. By 2020, smartphones reached over 60\% of the global population (with more than 80\% penetration in developed regions). The adoption curve was steeper than both PCs and the internet, reflecting the dramatic reduction in device costs and the integration of multiple functions (communication, navigation, entertainment, and productivity) into a single pocket-sized device \citep{west2012mobile}.

\item Artificial Intelligence Revolution (2020s -- Ongoing):
The current revolution, driven by artificial intelligence, marks a further leap. AI automates not only information retrieval but also knowledge processing: generating content, summarizing data, extracting insights, and even making decisions. Unlike earlier revolutions, AI democratizes not just access to tools but the act of cognition itself, enabling non-experts to harness systems that perform reasoning-like tasks \citep{brynjolfsson2023generative}.

The adoption speed is unprecedented. Within one to two years of release, generative AI systems reached hundreds of millions of users, with billions engaging indirectly through integrations into search engines, office software, and consumer applications. This scale and velocity of adoption make AI the fastest-diffusing technology in human history.
\end{itemize}

\subsection{A Four-Dimensional Perspective}


These revolutions can be mapped to a \emph{four-dimensional design space} :
(1) the human task automated at industrial scale;
(2) the \emph{aspect} of democratization (who gains access, R\&D, production, or usage);
(3) the \emph{degree} of democratization (how widely that capability is available);
and (4) the \emph{tempo of diffusion} (adoption speed and penetration over time). These four dimensions are described in Table \ref{tab2x}.



Each new wave extends all axes, tackling more complex tasks while speeding up broad access. Across five decades, the pattern is unmistakable: capabilities once confined to experts and institutions have become common tools. Given its scale and velocity, the AI wave may be the most transformative phase yet.

\begin{table}[h!]
\centering

\caption{Comparison of Four Technological Revolutions (1970s--2020s)}
\label{tab2x}
\renewcommand{\arraystretch}{1.3}
\begin{tabular}
{|p{4cm}|p{3.5cm}|p{3.5cm}|p{3.5cm}|}
\hline
\textbf{Late 1970s--Early 1980s} & \textbf{Late 1980s--Early 1990s} & \textbf{Late 1990s--Early 2000s} & \textbf{2020s--Ongoing} \\
\hline
\textbf{Personal Computing} & \textbf{Internet} & \textbf{Mobile \& Smartphones} & \textbf{Artificial Intelligence (AI Revolution)} \\
\hline
\textbf{Automated:} Data Processing & 
\textbf{Automated:} Information Sharing \& Access & 
\textbf{Automated:} Personal Communication \& Mobile Data Access & 
\textbf{Automating:} Knowledge Processing \& Decision Making \\
\hline
\textbf{Democratized:} Computing Power &
\textbf{Democratized:} Information Access &
\textbf{Democratized:} Instant Connectivity \& Information Creation &
\textbf{Democratizing:} AI Usage (Fastest in history) \\
\hline
\textbf{Penetration Rate:} $\sim$10\% by mid-1990s (mostly developed countries) &
\textbf{Penetration Rate:} $\sim$50\% by 2018 (25 years to reach 50\%) &
\textbf{Penetration Rate:} $>$60\% globally by 2020; $>$80\% in developed countries &
\textbf{Penetration Rate:} Hundreds of millions within 1--2 years; billions via integrations \\
\hline
\textbf{Adoption Speed:} Slowest &
\textbf{Adoption Speed:} Faster than PCs &
\textbf{Adoption Speed:} Faster than Internet &
\textbf{Adoption Speed:} Fastest in history \\
\hline
\end{tabular}
\end{table}

\begin{figure}[h!]
\centering
\includegraphics[width=0.8\textwidth]{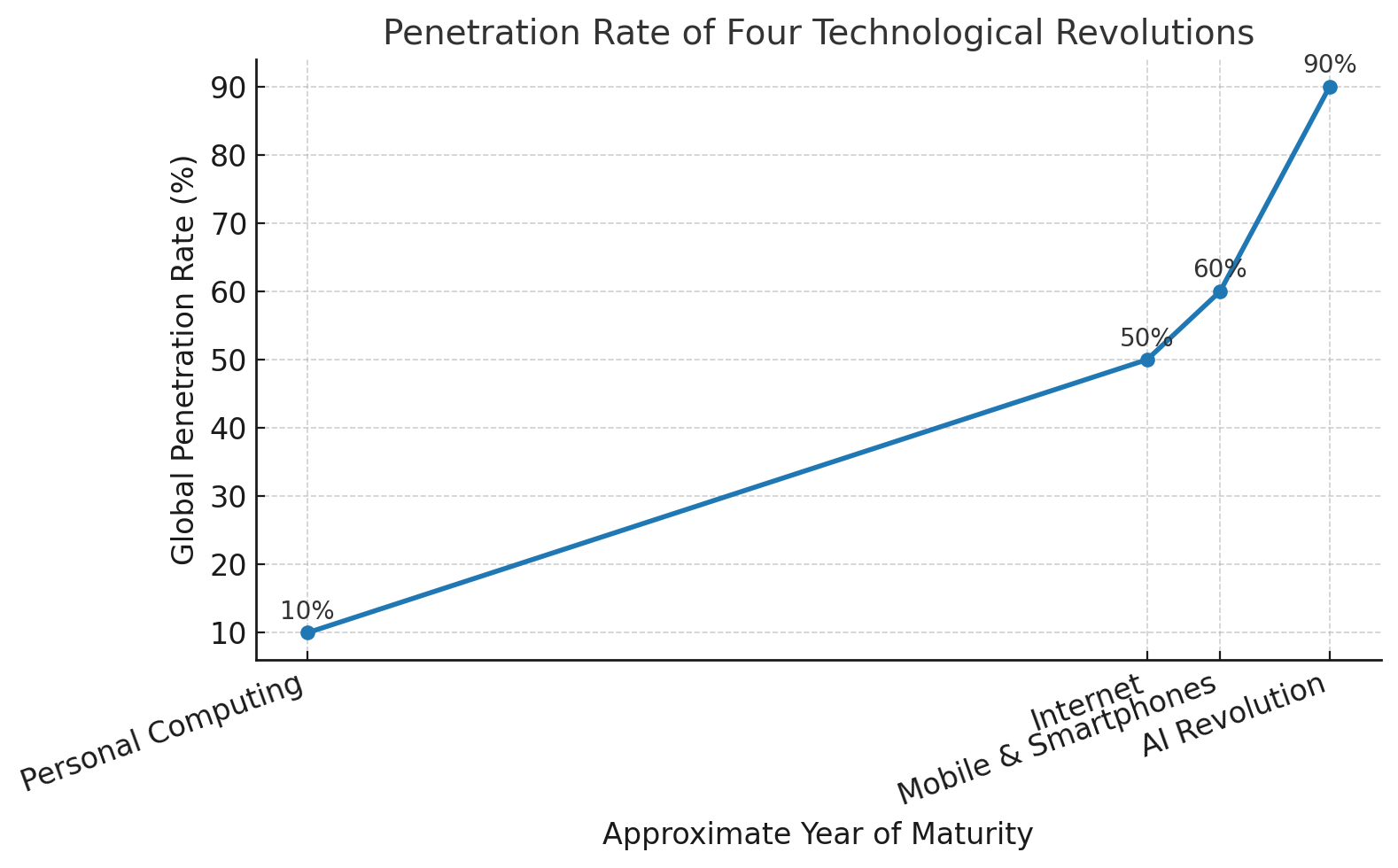}
\caption{Penetration rate of four major technological revolutions: Personal Computing, Internet, Mobile \& Smartphones, and Artificial Intelligence. The figure illustrates the accelerating pace of global adoption, with AI showing the fastest and broadest diffusion curve in history.}
\label{fig:tech_penetration}
\end{figure}

As shown in Figure~\ref{fig:tech_penetration}, each successive technological revolution has reached global penetration at an increasingly rapid pace. 
While personal computing took nearly two decades to reach 10\% penetration, the Internet achieved 50\% in roughly the same period, and mobile technologies surpassed 60\% by 2020. 
In contrast, the AI revolution has already reached hundreds of millions of users within just a few years, underscoring an unprecedented rate of adoption.

\subsection{Common Patterns Across Technological Revolutions}

A review of these four revolutions reveals several recurring patterns:
\begin{itemize}
    \item Democratization happens in layers: Broad democratization consistently occurs at the usage layer rather than production or R\&D. While many people can use tools, far fewer can build them, and only a small fraction push the research frontier. In PCs, Microsoft and Apple consolidated platforms; in the internet, Google and a few others dominated; in mobile, Android and iOS formed a duopoly. AI appears to be repeating this pattern, with a handful of frontier-model providers capturing most of the market, while open-source projects provide a vibrant countercurrent.

\item Market shares follow a power law: Each revolution produces a small number of dominant firms alongside a long tail of niche providers. This Zipf-like distribution can be seen in operating systems, search engines, app stores, cloud providers, and is now visible in AI ecosystems.

\item Open source is the durable counterweight: In every wave, open-source movements have pushed democratization deeper into production and R\&D. Linux shaped servers, TCP/IP and HTML sustained the web, and open-weight AI models enable local and private deployments. Open source remains a structural alternative to commercial monopolies.

\item Scale is defining: Each revolution multiplies access and throughput. PCs scaled personal compute cycles; the internet scaled global information; mobile scaled always-on connectivity; AI scales knowledge processing and decision assistance at unprecedented levels.

\item Costs decline sharply: Each wave witnesses dramatic reductions in unit costs (per FLOP, per byte, per inference), enabling new products and reinforcing adoption.

\item Personalization becomes the interface: Personalization runs through all revolutions: PCs individualized computing, the internet personalized content, mobile personalized connectivity, and AI personalizes cognition through adaptive co-pilots.

\item Each wave builds on the previous: The internet leveraged PCs; mobile leveraged the internet; AI leverages all three plus modern cloud and data pipelines.

\item Privacy is traded for convenience: Every revolution has deepened the trade-off between privacy and convenience. AI intensifies this tension, as models infer sensitive attributes from seemingly innocuous traces.

\item Inequality and moral distance widen: Each wave risks amplifying inequalities. In AI, access to high-end models and compute could deepen opportunity gaps and widen moral distance---the felt separation between one’s actions and their societal consequences \citep{zuboff2019age}.

\item Personalization is also an attack surface: Personalized systems reduce friction but increase exposure to tailored phishing, biased feedback loops, and over-delegation risks.
\end{itemize}
\paragraph*{}

The implications for AI are direct:
\begin{itemize}
    \item Expect consolidation at core layers (foundation models, orchestration platforms, app distribution), tempered by open-source and regulatory pressure.
    \item Push democratization upstream by investing in open models, transparent evaluation, affordable compute, and AI education.
    \item Design for privacy by default through on-device inference, data minimization, and auditable processes.
    \item Guard against widening moral distance by broadening access to high-quality AI and aligning incentives beyond commercial engagement.
    \item Treat personalization as critical infrastructure: secure it, test it, and empower users with control over data and agent autonomy.
\end{itemize}

In short, while the surface technologies have changed---from PCs to the web, to mobile, to AI---the structural patterns repeat. Recognizing these patterns allows societies to design AI systems that capture benefits while limiting harms.

\section{Changing How We See}

The transformative power of artificial intelligence does not end with altering modes of production. Like industrialization, it eventually reshapes how people perceive, interpret, and interact with the world. In the nineteenth century, industrialization standardized products, distribution channels, and consumer expectations. Similarly, AI---particularly large language models---is changing not only the content we produce but also the way we value, consume, and trust knowledge.

\subsection{From Production to Perception}

During industrialization, automation first changed how goods were produced. To reduce costs and increase efficiency, producers shifted from artisanal, variable goods to standardized mass products. Over time, this standardization changed consumption itself: buyers came to expect uniformity, reliability, and low prices. In parallel, AI is transforming knowledge production. Content that once required hours of human effort can now be generated in seconds. The result is not merely more content but content that is increasingly standardized in style, structure, and reliability \citep{beniger1986control}.

This shift extends to perception. Just as industrial production altered how consumers judged value in material goods, AI will change how we assess and interpret information. In a world of machine-generated outputs, human judgment migrates from production to verification, curation, and contextualization.

\subsection{The Three-Stage Content Economy}

The AI content economy can be mapped in three stages:
\begin{itemize}
    \item Stage 1 --- Production Shift: 
Content creation moves from artisanal production to automated pipelines. With the marginal cost of producing drafts approaching zero, teams can generate dozens of iterations before selecting a final version. Style guides and prompts increasingly function as assembly-line instructions.

\item Stage 2 --- Product Shift: 
As AI-assisted content becomes the default for routine communication, the market polarizes. Commodity outputs such as emails, lesson plans, or basic reports become nearly free. Meanwhile, ultra-bespoke or highly trusted human work becomes more valuable, commanding premium prices. This bifurcation mirrors the industrial “barbell effect.”

\item Stage 3 --- Perception Shift:
The deepest transformation occurs in how we read and interpret content. Readers adopt a “verification-first” stance, expecting provenance, citations, and uncertainty markers. Research shifts from source-hunting to orchestrating multiple AI-backed perspectives. Writers become system designers, specifying parameters and verifying outputs rather than crafting every sentence. Human creativity moves from typing to framing, editing, and imbuing content with ethical and contextual depth.
\end{itemize}

The vase parable illustrates this arc. In the handcrafted era, vases varied by style and quality, with the artisan’s name serving as a guarantee. Industrialization brought uniform, machine-made vases that initially signaled progress and aspiration. Eventually, however, uniqueness regained value: bespoke, signed pieces became luxury goods. AI follows the same trajectory. Low-tier, formulaic writing will vanish as machines outperform humans at speed and cost. High-value work will concentrate where human judgment, authenticity, and originality cannot be replicated \citep{walsh2023machines}.


Variance is expensive, whether in manufacturing or information systems. Industrial production punished variance because it increased costs. In AI, variance shows up as inconsistency, compliance risk, or review burden. Consequently, prompts, templates, and checklists increasingly function as the “rails” of cognition. Just as assembly lines structured physical production, interaction patterns with AI systems will structure intellectual work.

\subsection{Implications for Knowledge and Culture}

AI is changing not only how knowledge is produced but also how it is trusted and consumed. Readers and institutions must adapt to a landscape in which:
\begin{itemize}
    \item Commodity content proliferates, while authenticity and provenance become premium signals.
    \item Educational systems must train students in verification-first reading and system-level writing, not only prose generation.
    \item Institutions must design review pipelines---bias checks, fact-checking, red-teaming---to standardize quality in ways analogous to industrial quality control.
    \item Cultural norms may shift toward valuing distinctiveness, accountability, and transparency as markers of human authorship.
\end{itemize}


AI’s impact extends beyond “changing the look” of outputs. It transforms how we see, read, and interpret information. In this transition, the locus of human value shifts to judgment, taste, ethical responsibility, and contextual framing. Preparing for this perceptual shift is as important as preparing for the productivity gains of automation.

\section{Implications for How We Perceive Text}

Artificial intelligence not only accelerates the production of written material but also reshapes what text \textit{is}, how it is consumed, and how it functions within society. Media theorists have long emphasized that the medium is not neutral: the tools we use to produce communication influence style, tone, and even what counts as knowledge \citep{mcluhan1964understanding}. Just as Nietzsche remarked that his prose changed when he adopted the typewriter, we must consider how word processors, grammar checkers, and now large language models alter both writing and reading practices.

\subsection{From Static Artifact to Interactive Partner}

Traditionally, a text has been treated as a static artifact: a book, article, or essay fixed in form. AI enables a transition toward interactive, adaptive documents. Readers will increasingly expect to “talk” to texts---asking questions, testing claims, or requesting counterexamples within the document itself. A policy report, for example, might allow readers to adjust assumptions and immediately regenerate projections. This turns the text into a \textit{system} rather than a monologue.

Moreover, texts may become multimodal by default. A passage could spawn a diagram, simulation, or audio gloss on demand. In this sense, the written word becomes the entry point into an interactive knowledge environment rather than the endpoint of expression.

\subsection{How AI Changes Writing (Supply Side)}

For writers, the supply-side implications are profound. New capabilities include instant outlining, multilingual drafting, style transfers, data-aware passages, and auto-fact checking. These reduce friction and broaden access. At the same time, there is a tendency toward homogenization: without deliberate effort, AI-generated text converges toward “average style.” Distinctive authorial voice becomes a scarce asset.

Authorial labor also shifts. Writers spend less time typing and more time designing prompts, specifying constraints, and curating tone. Verification of truth and ethical alignment become as central as expression itself. In effect, authors move from wordsmiths to system designers of textual production \citep{vincent2023writing}.

\subsection{How AI Changes Reading (Demand Side)}

On the demand side, expectations rise. Readers will come to expect provenance, citations, and uncertainty cues as defaults. Instead of consuming and trusting, readers will probe, verify, and then use information. Adaptive interfaces will allow users to request summaries for novices, expert-level elaborations, or translations into other languages. Attention will shift away from decoding prose toward evaluating quality, ethics, and consequences.

\subsection{Structural Shifts in the Meaning of Text}

AI induces several structural transformations in how text is defined:
\begin{itemize}
    \item From linear to branching: Readers navigate through interactive question paths rather than fixed pages. 
    \item From monologue to dialogue: Documents can respond and remember conversational context. 
    \item From monolingual to polyglot: Seamless translation preserves tone and idiom, eroding linguistic barriers. 
    \item From opaque to auditable: Embedded citations, data lineage, and uncertainty notes become standard. 
    \item From one-size-fits-all to adaptive: Documents shift tone, difficulty, and examples based on reader profiles. 
\end{itemize}


These dynamics raise the possibility that static, non-interactive text may come to feel obsolete in technical, educational, and policy domains. For legal documents, technical manuals, and instructional content, the absence of interactivity or verification could be seen as a deficiency. However, for literary art---poetry, novels, and essays---the scarcity of human voice and intentionality remains central. Here, interactivity may enrich discovery but is not necessary for value.


For authors, editors, and educators, several practices emerge:
\begin{itemize}
    \item Writers should embed provenance, source documentation, and uncertainty markers into their work. 
    \item Editors and publishers should ship documents with conversational layers, facts/evidence panels, and audit trails of what was human versus machine-generated. 
    \item Educators should emphasize dialogic reading and system-level writing, training students to interrogate, verify, and evaluate rather than merely reproduce. 
\end{itemize}


Text is evolving from a fixed artifact into a dynamic, interactive system. Readers will expect conversation, provenance, and personalization by default. In technical and policy domains, non-interactive text may feel obsolete; in literature, voice and authenticity remain irreplaceable. The deeper challenge lies in ensuring that as AI transforms language, humans retain judgment, accountability, and creativity as central to meaning-making.

\section{Conclusion}

The trajectory of artificial intelligence can be understood only by holding multiple perspectives in tension. Through the risk lens, AI resembles nuclear technologies: it carries irreversible and potentially catastrophic tail risks that demand global governance at software speed. Through the transformation lens, it mirrors the Industrial Revolution: a general-purpose technology that reorganizes economies, reshapes labor, and produces turbulence as societies adapt. Through the continuity lens, it extends a fifty-year arc of computing revolutions---from personal computers to the internet to mobile---exhibiting recurring structural patterns of democratization at the user layer, concentration in production, falling costs, and rising personalization.

Historical analogies remind us that no true singularities exist. Past technological transitions---from writing and the printing press to industrialization and the internet---were disruptive, yet they ultimately became governable as new norms, standards, and institutions emerged. The closest parallels for AI lie in the Industrial Revolution, which broke economic growth ceilings, and nuclear weapons, which introduced civilization-scale risks. These episodes suggest a dual posture: treat AI as governable in its median effects while preparing robust guardrails for its tail risks.

The near-term future points toward the deflation of cognitive services, reshaping industries from law and education to translation and software engineering. Routine reasoning will be commoditized, while scarcity will migrate to judgment, trust, and ethical responsibility. At the same time, adoption will amplify familiar tensions: privacy traded for convenience, inequality deepened by asymmetric access, and moral distance widened as decision-making becomes mediated by algorithms.

The design of moral AI agents represents the frontier challenge. Ensuring that autonomous systems remain aligned with human values requires integrating specification, verification, and enforcement mechanisms, fostering moral generalization in open environments, and governing emergent dynamics in multi-agent ecosystems. This endeavor demands interdisciplinary collaboration: computer scientists to engineer reliability, philosophers and social scientists to articulate values, and policymakers to construct accountability regimes.

In sum, AI is neither wholly revolutionary nor merely evolutionary. It is both continuity and rupture: familiar enough to be governed with known tools, yet powerful enough to warrant vigilance. The outcome will depend not on algorithms alone but on the institutions, safeguards, and cultural norms we build around them. By recognizing AI as mathematics and infrastructure---not magic---and by embedding it in a human order of responsibility, societies can capture its transformative benefits while containing its existential risks.

\bibliographystyle{plainnat}

\end{document}